# Smartphone App to Investigate the Relationship between Social Connectivity and Mental Health*

Tjeerd W. Boonstra, Aliza Werner-Seidler, Bridianne O'Dea, Mark E. Larsen, and Helen Christensen

*Abstract*— Interpersonal relationships are necessary for successful daily functioning and wellbeing. Numerous studies have demonstrated the importance of social connectivity for mental health, both through direct peer-to-peer influence and by the location of individuals within their social network. Passive monitoring using smartphones provides an advanced tool to map social networks based on the proximity between individuals. This study investigates the feasibility of using a smartphone app to measure and assess the relationship between social network metrics and mental health. The app collected Bluetooth and mental health data in 63 participants. Social networks of proximity were estimated from Bluetooth data and 95% of the edges were scanned at least every 30 minutes. The majority of participants found this method of data collection acceptable and reported that they would be likely to participate in future studies using this app. These findings demonstrate the feasibility of using a smartphone app that participants can install on their own phone to investigate the relationship between social connectivity and mental health.

## I. Introduction

Social connections are critically important to health and wellbeing, with numerous studies finding that aspects of social connection affect morbidity and mortality at levels comparable to smoking and obesity [1]. Social connectivity also shares an important and specific relationship with mental health, with studies showing that individuals who have smaller networks, fewer interpersonal relationships or lower levels of social support consistently report elevated rates of depression [2, 3]. Furthermore, a significant proportion of those who die by suicide have a history of social isolation [4].

Although both prospective and cross-sectional studies have been conducted to investigate the association between social network factors and mental illness, these studies rely primarily on questionnaire data. Studies have typically depended on an array of self-report indices, such as name generators (identifying who is in one's social network), number of friends, frequency of participation in social activity, and whether someone is living alone or not, in order to identify the size and nature of social networks [5, 6]. However, these measures are limited by the nature of self-report – which is inherently vulnerable to confounding and systematic bias [7]. These methods are also practically limited by the time and effort required to administer and process such data. This can severely impede the extent to which certain populations, communities, and individual networks can be studied. New technologies are needed to advance the field of social network analysis.

Sensor-enabled mobile phones have recently been tested as a method to measure the proximity between participants and map face-to-face interactions, with evidence showing that proximity is a valid metric of social connections [8]. Social connections can be accurately deduced from proximity by distinguishing typical proximal behavioral patterns (for example, proximity to colleagues at work during the day) from other behaviors (proximity to others outside of workplace). However, this technology has been limited by several factors. First, Bluetooth data has only been captured on Nokia smartphones [8] or those with an Android operating system [9]. As such, this technology is not available to the estimated 700 million iPhone users who rely on the iOS operating system. Second, past studies involved pre-installing the application on a device that was then distributed to participants for the duration of the study [10, 11]. This limits not only the generalizability of findings, but also the ability to capitalize upon and use technologies at scale.

The current study builds upon existing research to test the feasibility of a smartphone application (app) developed by our team for the objective measurement of an individual's social network. Participants can install the app on their own smartphone running either an Android or iOS operating system. In this study we examined whether it is feasible to use this technology to collect Bluetooth and mental health data in a naturalistic setting.

## II. Method and Procedures

### A. Participants and procedure

The app was tested in a work environment where the boundaries of the social network can be easily defined. Staff from an international data analytics company located in Sydney was invited to join the study via an email that contained the participant information sheet. Participants were required to be in the Sydney office for at least two days each week over the four-week study period (28 March – 24 April 2016). If they were interested in participating, they could register on the study website. They then received an email containing a URL to install the app on their phone. The consent form was embedded within the app and participants were prompted to provide electronic consent when they opened the app for the first time. This study was approved by the University of New South Wales Human Research Ethics Committee (HC15526).

* Research supported by the NHMRC Centre of Research Excellence in Suicide Prevention APP1042580 and NHMRC John Cade Fellowship APP1056964. B. O'Dea and M. E. Larsen are supported by Society of Mental Health 2015 Early Career Research Awards.

T. W. Boonstra, A. Werner-Seidler, B. O'Dea, M. E. Larsen, H. Christensen are with the Black Dog Institute, University of New South Wales, Sydney, NSW 2031, Australia. T. W. Boonstra is also with QIMR Berghofer Medial Research Institute, Brisbane, Australia. Email: t.boonstra@unsw.edu, {a.werner-seidler, b.odea, mark.larsen, h.christensen}@blackdog.org.au.

## B  Data acquisition using smartphone app

All data in the study were collected using a custom-built smartphone application developed by the research team for both Android and iOS operating systems [12]. The Research Kit framework for iOS [13] was used to generate the informed consent form and surveys within the app, and this functionality was replicated on Android.

After completion of the consent form, participants were asked to complete several psychometrically validated mental health questionnaires also embedded in the app, including the Patient Health Questionnaire (PHQ-9, depression) [14] and Generalized Anxiety Disorder Scale (GAD-7, anxiety) [15].

Subsequently, Bluetooth data were passively collected for four weeks. For the iOS application, the *BluetoothManager* private API was used, as the public *CoreBluetooth* API only contains functions for interacting with low-energy devices and it is currently not feasible to use Bluetooth Low Energy to map social networks in iOS [16]. The app was configured to perform a Bluetooth discovery scan every five minutes during the study period. When a Bluetooth device conducts a discovery scan, other Bluetooth devices within a range of 5–10 m respond with their user-defined name, the device type, and a unique 12-hexadecimal-digit hardware media access control (MAC) address. When a participant's MAC address is discovered by a periodic Bluetooth scan performed by another participant, it indicates that both smartphones are within 5–10 m of each other (see also [12]). These data were cryptographically hashed on the handset to ensure the privacy of participants.

At the end of study, participants completed an exit survey assessing user experience including installation problems, battery usage, and the perceived impact on their privacy.

## D. Data analysis

Although the app detects any Bluetooth device that is in close proximity, in the current study we only analysed the connections between participants. The social proximity network was estimated based on the Bluetooth scanning statistics of participants' smartphones. The connection strengths between participants (edge weights) were defined as

$$R_{ij} = \frac{N_{ij}(T)+N_{ji}(T)}{N_i(T)+N_j(T)}, \quad (1)$$

where $N_{ij}$ is the number of scans where device *i* detected device *j* and $N_i$ the number of times device *i* scanned on time interval *T* [12]. By normalizing the number of times one device detected the other by the number of times each device scanned, the connection strength $R_{ij}$ is bound on the interval [0,1], where 1 indicates that both devices always detected each other when they scanned and 0 indicates that the devices never detected each other.

Weighted networks were constructed based on Bluetooth data aggregated across the 4-week study period. A binary backbone network was extracted from the weighted networks obtained using the app [17]. This filtering method provides a statistical method to extract the relevant connection backbone in complex networks by preserving edges that are statistically significant. The R package 'disparityfilter' was used to extract the backbone network [18]. Network topology was generated using the Fruchterman-Reingold algorithm and PHQ-9 scores were used as node characteristic.

## III.  RESULTS

Of the 74 people that registered on the study website, 70 installed the study app on their smartphone. Of these 70 participants, 63 completed the study: three did not work in the Sydney office, one participant had problems installing the app, one participant had a smartphone with OS that was too old (iOS 7), and two participants withdrew from study. Hence, the drop-out rate was 11/74 = 15%. Of the 63 participants who completed the study, 7 used Android and 56 used iOS. The participants worked across different departments, including Sales, Administration, Technical Support, Education and Finance.

### A. Mental health scores

56 participants completed the mental health questionnaires (PHQ-9 and GAD-7). Based on the PHQ-9 scores, 38 participants (68%) had minimal depression, 12 (21%) mild, four (7%) moderate, none had moderately-severe, and two (4%) had severe depression [14]. Based on the GAD-7 scores, 44 participants (79%) had minimal anxiety, 10 (18%) mild, one (2%) moderate, and one (2%) severe anxiety [15]. Fig. 1 shows the distribution of PHQ-9 and GAD-7 scores.

### B. Scanning rates

Bluetooth data were obtained from 54 participants. Scanning rates varied considerably across participants and were generally higher on devices running Android compared to iOS (Fig. 2A). On average 32% of the scheduled scans were performed (iOS: 29%, Android: 55%). To estimate edge weights between participants the scanning statistics of each pair of smartphones were combining (Eq. 1). The scanning rates for the edges were therefore higher than for individual devices (Fig. 2B).

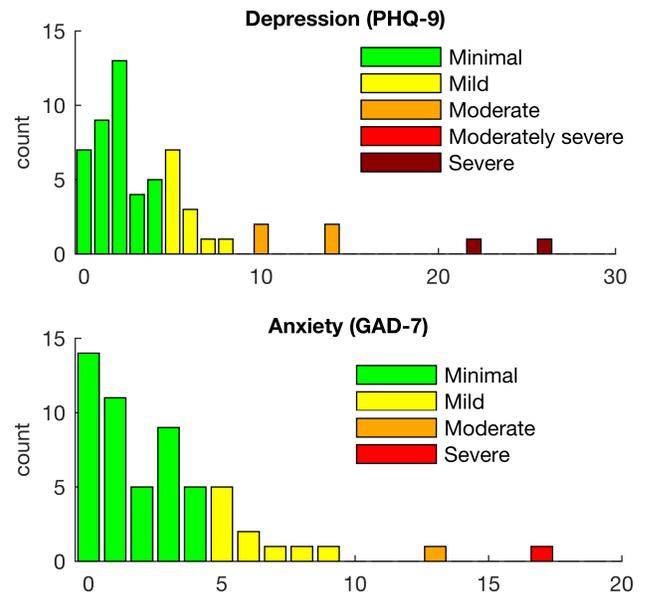

Figure 1.  Mental health scores. Top panel shows the distribution of PHQ-9 scores (depression) and the bottom panel the GAD-7 scores (anxiety).

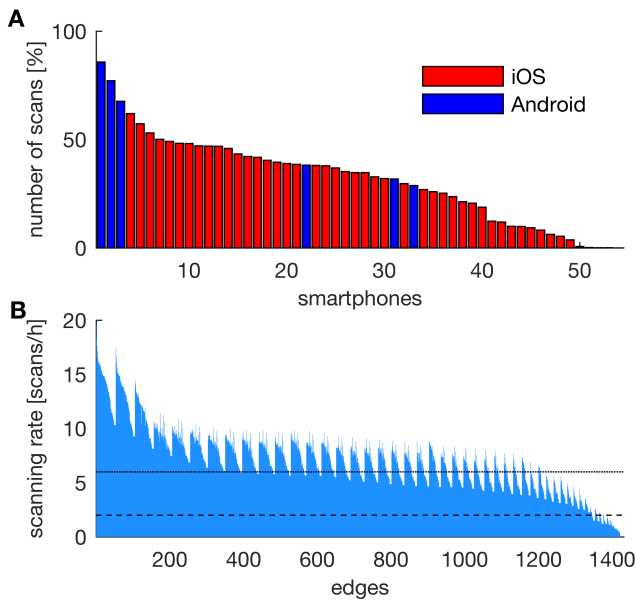

Figure 2. Bluetooth scanning rates. Top panel shows the percentage of scheduled scans that were performed on each device and bottom panel the average scanning rate for each edge of the social network.

Of all 1431 edges, 62.8% of the edges were scanned al least every 10 minutes and 94.7% of the edges were scanned at least every 30 minutes.

*C. Social network of proximity*

Networks were generated by aggregating data across the 4-week study period and preserving edges that are statistically significant. The network reveals which participants were in close proximity during the study. The PHQ-9 scores of the participants were used to color-code the nodes of the social network to investigate the relationship between network topology and mental health (Fig. 3).

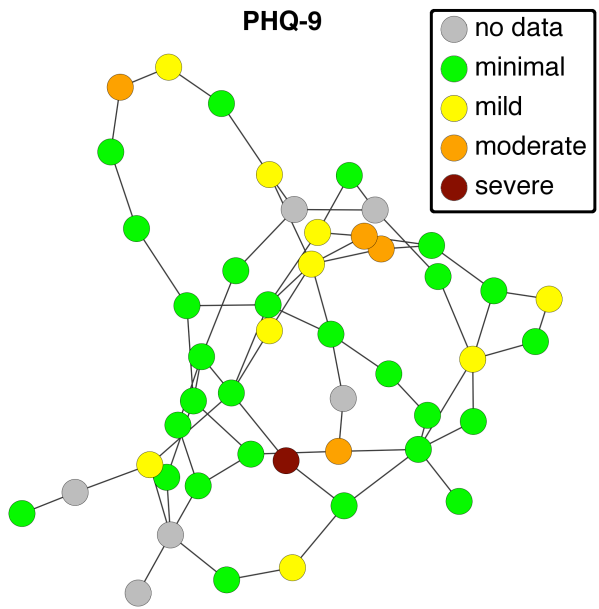

Figure 3. Social network of proximity. Proximity was estimated from Bluetooth data. Signficant edges were estimated using disparityfilter. Node color indicate the depression score of participants based on the PHQ-9.

Although the distribution of PHQ-9 scores are skewed towards zero, thus making it difficult to interpret meaningful patterns, by inspection it appears that there may be some degree of clustering of participants with moderate or severe PHQ-9 scores. Future studies are needed, with larger sample sizes and a broader range of symptom scores, to allow further analysis of clustering and centrality within the network.

*D. Exit survey*

In the exit survey, one participant (2%) reported difficulties installing the app and two participants (4%) reported issues having the app run in the background. The user experience was overall positive. Most participants (50 %) reported that the app did not impact on the battery life of the smartphone at all or very little, although four participants (8%) thought the app impacted battery life very much (Fig. 4). The majority of participants (50%) reported that the app did not impact on their privacy at all and 21 participants (42%) said it was very likely that they would participate in a similar study using this app again.

## IV. DISCUSSION

The results of this study show that it is possible to obtain Bluetooth sensor data from participants' own smartphones for the purposes of mapping social networks based on proximity. To our knowledge, this is the first study to demonstrate the ability to collect meaningful Bluetooth data on a range of different smartphone types, across both Android and iOS operating systems.

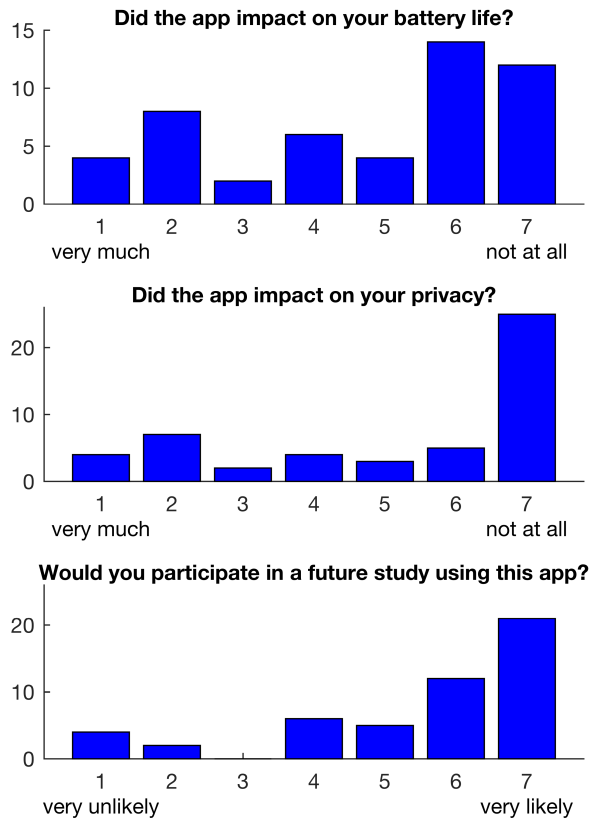

Figure 4. User experience. Top panel shows perceived impact on battery life, middle panel perceived impact on privacy and bottom panel willingness to participate in future studies using this app.

Moreover, this study has demonstrated that a group of unselected individuals find this method of data collection not only acceptable (15% drop-out rate), but a significant proportion of participants reported that they would be very likely to participate in future studies using this app. Finally, most participant thought the app did not impact on their privacy or battery use, suggesting that this is both a feasible and acceptable method of data collection. That said, a few participants reported that battery life was an issue for them. It will be important in future studies of this nature to assess phone age and type in order to identify which battery types are likely to be impacted.

The study did find however, inconsistent scanning patterns for Bluetooth data. There were several factors that may have reduced the completeness of the Bluetooth data which warrant discussion. First, scanning rates were superior on Android relative to iOS. This is likely because of restrictions in iOS to apps running in the background. Second, as is inherently the case when conducting experimental studies in real-world naturalistic settings, scanning rates were likely to be influenced by individual participant behavior. In previous studies participants were provided with phones that already had the application installed and were configured to allow Bluetooth data to be collected. This means that all the necessary permissions and notifications had already been enabled. Disabling such functions would require the user to proactively switch them off. This is a conflicting issue as the benefit in using participants' own smartphones means that this method is scalable; however, it also means that users have much greater control over the settings and functions of their smartphones. For example, participants may have turned off their Bluetooth data during the study, while others may have lost internet connection or used up all of their data allowance, all of which would have reduced the completeness of data obtained using the app. The next step in this line of enquiry is to systematically test the impact of different user behaviors on scanning rates, as well as address technological factors that may have limited the completeness of the data. This work is currently underway in our laboratory.

This study demonstrates that Bluetooth data collected on participants' own smartphone can be used to map social networks and that people are prepared to provide smartphone data and mental health data, increasing the feasibility of undertaking this potentially sensitive research to a larger scale. Social network factors are known to be critically important for mental health, and studies involving a larger sample size are needed to reliably identify social network features that are predictive of mental illness such as depression, anxiety and suicidality [19]. This kind of objective information has the capacity to improve the detection of mental illness using objective social network indices [20], provide risk factors for suicidality by detecting social withdrawal [21], as well as potentially suggest novel avenues for intervention. If this method of passive data collection reported here can be replicated and scaled, it has the potential to represent a methodological paradigm shift in the field of mental health and social connectivity given the historical reliance on self-report methods that are subject to bias.

ACKNOWLEDGMENT

We thank Samuel Townsend for providing technical support during the study.